# BanglaShop-CRS: A User-Centric Bangla Dataset for Conversational Recommendation

**Tabia Tanzin Prama**[1,2*], **Christopher M. Danforth**[1,3], **Peter Sheridan Dodds**[1,2,4,5†]

[1] Computational Story Lab, Vermont Complex Systems Institute, MassMutual Center of Excellence for Complex Systems and Data Science, Vermont Advanced Computing Center, University of Vermont, Burlington, VT 05405, US
[2] Department of Computer Science, University of Vermont, Burlington, VT 05405, US
[3] Department of Mathematics & Statistics, University of Vermont, Burlington, VT 05405, US
[4] Santa Fe Institute, 1399 Hyde Park Rd, Santa Fe, NM 87501, US
[5] Complexity Science Hub, Metternichgasse 8, 1030 Vienna, Austria

**Logline**

BanglaShop-CRS introduces a large-scale synthetic Bangla conversational recommendation dataset grounded in real e-commerce user behavior to support research for low-resource Conversational recommender systems (CRS).

**User:** আসসালামু আলাইকুম! আমি ধর্মীয় সামগ্রী নিয়ে কথা বলতে চাই।

**Agent:** আলহামদুলিল্লাহ, কীভাবে সাহায্য করতে পারি?

**User:** Storage & Organisation নিয়ে কোনো সুপারিশ করবেন?

**Agent:** Storage & Organisation নিয়ে আমি এ ধরনের বিকল্প সুপারিশ করি: প্লাস্টিক স্টোরেজ বক্স, ড্রয়ার অর্গ্যানাইজার, ওয়াল-স্টোরেজ সলিউশন। এগুলো ধর্মীয় সামগ্রী আলাদা করে সাজাতে সাহায্য করবে।

**User:** ঠিক আছে, এ সুপারিশ গ্রহণ করছি। Agent: ধন্যবাদ কথা বলার জন্য। যদি আরও সাহায্য চান, বলবেন। শুভ দিন!

*tabia.prama@uvm.edu
†peter.dodds@uvm.edu

## Abstract

Conversational recommender systems (CRS) enable users to express preferences, constraints, and feedback through natural language interaction. However, existing CRS resources are concentrated in English and other high-resource languages, leaving Bangla and code-mixed Bangla--English settings underrepresented. To address this gap, we introduce BanglaShop-CRS, a large-scale user-centric synthetic Bangla conversational recommendation dataset grounded in real e-commerce behavior. It contains 27,178 multi-turn dialogues, 274,802 utterances, and 3.6M tokens across 10 product domains. Our generation pipeline incorporates user purchase histories, positive and negative feedback, and review texts to maintain consistency between dialogue content and user preferences. We evaluate BanglaShop-CRS under catalog-constrained and open-vocabulary recommendation protocols. Results show that dialogue context improves recommendation quality, while fine-tuning yields further gains across models. Human evaluation by five native Bangla-speaking annotators confirms the fluency, informativeness, logicality, and coherence of the dialogues, with significant agreement across all dimensions. Factual-grounding evaluation further shows stronger alignment with correct than shuffled user records, with substantial inter-annotator agreement ($\kappa = 0.65$) and comparable human and GPT-5.1 judgments. BanglaShop-CRS provides a scalable benchmark for advancing conversational recommendation in Bangla.

## Contents

# 1 Introduction

Recent advances in generative artificial intelligence (GenAI) have reshaped human-computer interaction (HCI), largely through large language models (LLMs) that can produce coherent, context-aware, and personalized natural language responses [1, 2]. A key application of this progress is conversational recommender systems (CRS), which provide personalized recommendations through natural language dialogue [3--5]. Unlike traditional recommender systems, which mainly rely on user-item interaction histories [6], CRS allow users to express preferences, constraints, and feedback through multi-turn conversations. This makes the recommendation process more interactive, adaptive, and user-centered [7]. Despite these advances, effective CRS development still depends on the availability of large-scale, high-quality conversational datasets. Existing benchmarks such as REDIAL [8], TG-ReDial [9], and DuRecDial [10] have played an important role in advancing CRS research. REDIAL introduced over 10,000 human-human movie recommendation dialogues, while TG-ReDial and DuRecDial extended the field through topic-guided and goal-oriented dialogue construction. More recently, LLM-REDIAL demonstrated that LLMs can support scalable synthetic CRS dataset creation while preserving consistency between dialogue content and users' historical behaviors [11]. However, most existing CRS datasets are still concentrated in English and other high-resource languages, leaving low-resource languages substantially underrepresented [12].

This gap is particularly important for Bangla, one of the world's most widely spoken languages, with more than 300 million speakers across Bangladesh and parts of India [13, 14]. In everyday digital communication, Bangla speakers often use mixed linguistic forms such as Banglish, where Bangla and English appear within the same utterance [15]. Such code-mixed and culturally grounded language use creates challenges for GenAI [16] and CRS systems, which often perform poorly for languages, dialects, and writing styles that are underrepresented in training data [17, 18]. Consequently, Bangla-speaking users may face misinterpretation, lower recommendation quality, reduced usability, and diminished trust in conversational systems [19].

Two key challenges limit the construction of CRS datasets for low-resource languages. First, existing dataset construction methods often require extensive human annotation, making them costly, time-consuming, and difficult to scale [20]. Moreover, dialogues collected through crowd-workers or retrieval-based methods can vary in quality and may not fully reflect users' real preferences. Second, many existing datasets suffer from semantic inconsistency between generated dialogues and users' actual historical behaviors [21]. Since recommendation quality depends not only on fluent conversation but also on accurately modeling user preferences, a CRS dataset should align dialogue content with users' positive feedback, negative feedback, reviews, and prior interactions [22].To address these limitations, we introduce BanglaShop-CRS, a large-scale synthetic Bangla conversational recommendation dataset generated from real user behavior using LLMs, following LLM-REDIAL [11]. It contains 27K multi-turn dialogues, 274.8K utterances, and 3.6M tokens across 10 product domains. Our contributions are:

- We construct user-grounded dialogues from purchase histories, ratings, and reviews, enabling consistent recommendations and long-term preference modeling.
- We evaluate the dataset through LLM-based recommendation experiments and human assessments at utterance, conversation and dialogue levels, providing a scalable resource for Bangla and other low-resource communities.

# 2 Dataset Construction

As shown in Figure 1, BanglaShop-CRS is constructed through a three-stage pipeline: data preprocessing, template construction, and dialogue generation. First, raw Daraz product reviews are filtered, grouped, and split to build per-user interaction histories and target item lists. Next, communicative goals are defined and organized into structured multi-turn dialogue templates. Finally, an LLM generates complete Bangla recommendation dialogues by filling these templates with real user behaviors and review information.

## 2.1 Data Source.

To ground BanglaShop-CRS in authentic user behavior, we source all product reviews from BanglishRev [23], the largest e-commerce product review dataset for the Bengali-speaking population to date. BanglishRev comprises 1.74 million written reviews drawn from 3.2 million ratings across 128k products sold on online platforms targeting Bengali speakers. The reviews are written in Bengali, English, code-mixed Bengali--English, and Banglish (Bengali words transcribed using the English alphabet), making the dataset a realistic reflection of how Bangladeshi users communicate in digital commerce contexts. Each user's numerical ratings are used to infer item preferences, which are then combined with the corresponding review texts to construct dialogue prompts. In this way, every generated dialogue in BanglaShop-CRS is directly associated with one specific user's historical interaction record, forming a complete behavioral foundation for conversational recommendation.

## 2.2 Data Preprocessing.

To prepare the raw review data for dialogue generation, we apply several preprocessing steps to ensure data quality and suitability for conversational recommendation. We tokenize all review texts and remove irregular or non-word tokens. To provide sufficient behavioral context, we discard users and items with fewer than 100 interactions. To capture both acceptance and rejection scenarios, we retain positive and negative preferences: ratings of 4 or higher are grouped into a Likes collection, while ratings of 2 or lower are grouped into a Dislikes collection. Both are sorted chronologically to preserve interaction order. Finally, the last 5% of each user's positive interactions are held out as Possible Recommendations, from which ground-truth target items are selected during evaluation.

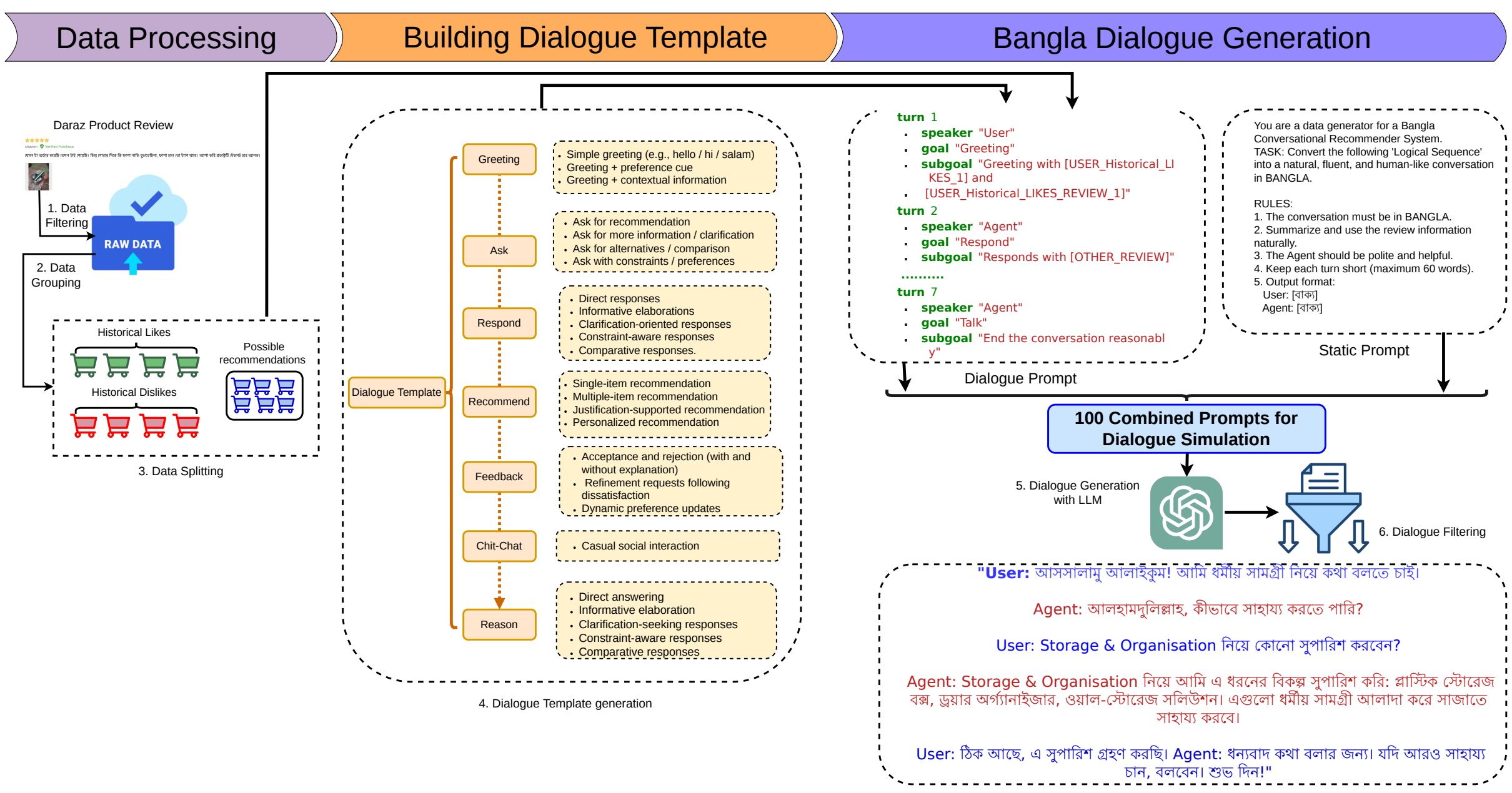

Figure 1: Overview of the BanglaShop-CRS dataset construction framework, consisting of three sequential stages: data preprocessing, template construction, and dialogue generation.

## 2.3 Template Construction

**Goal Design.** To guide structured recommendation-oriented dialogue generation, we define communicative goals for utterances following the dialogue-act taxonomy of ISO 24617-2. BanglaShop-CRS includes eight primary goals: *Greeting*, *Ask*, *Respond*, *Recommend*, *Feedback*, *Chit-Chat*, *Talk*, and *Reason*. Table 1 presents these primary goals and their descriptions. Each primary goal is further divided into fine-grained sub-goals, resulting in 25 unique sub-goals. These sub-goals define the specific communicative function of each utterance, such as initiating a conversation, requesting a recommendation, providing item information, or expressing preference feedback. The complete list of primary goals and sub-goals is provided in Appendix A1.2, Table A1.

Sub-goals are designed in two forms: fixed and slot-based. Fixed sub-goals provide static instructions, such as ``*Ask for recommendation*'' or ``*End the conversation reasonably*.'' Slot-based sub-goals contain placeholders filled with user-specific data, such as ``*Recommend [USER_MIGHT_LIKES]*'' or ``*Greeting with [HISTORICAL_LIKES] and [HISTORICAL_LIKES_REVIEW]*.'' These slots are populated using items and review texts from each user's Likes, Dislikes, and Possible Recommendations sets, ensuring that each generated dialogue remains grounded in user-specific behavioral evidence.

**Template Design.** To generate fluent and culturally grounded Bangla recommendation dialogues, we construct 100 dialogue templates. Each template defines a sequence of turns, where every turn is assigned a speaker role, primary goal, and sub-goal. The templates are grouped into three types based on the number of recommendation attempts: one, two, or three. In multi-recommendation templates, earlier suggestions may be rejected before the final recommendation is accepted, reflecting realistic conversational behavior. Dialogue lengths range from 6 to 16 turns, following common CRS benchmark patterns [8--10]. To preserve linguistic authenticity, the templates include Bangla and Banglish user reviews through slots such as `[USER_HIS_LIKES_REVIEW]`, `[USER_HIS_DISLIKES_REVIEW]`, and `[OTHER_MIGHTLIKES_REVIEW]`. Each prompt also instructs the LLM to generate the final dialogue in Bangla, ensuring both structured recommendation flow and Bangla-specific conversational style.

Table 1: Eight primary dialogue goals and their descriptions for utterances in the BanglaShop-CRS dataset.

| Primary Goal | Utterance Description |
|---|---|
| Greeting | Initiates the dialogue with a greeting, user preference cue, or relevant context. |
| Ask | Represents user requests for product recommendations, details, alternatives, or preferences. |
| Respond | Provides direct, informative, clarifying, constraint-aware, or comparative responses. |
| Recommend | Suggests products using personalization, justification, or evidence from user history. |
| Feedback | Captures user acceptance, rejection, dissatisfaction, or preferences after a suggestion. |
| Chit-Chat | Adds casual interaction to make the dialogue more natural and conversational. |
| Talk | Guides the conversation toward recommendation or closes it appropriately. |
| Reason | Explains user decisions or rejections using prior experience, constraints, or comparisons. |

# 3 Dialogue Generation

## 3.1 Generation with LLMs.

As shown in Figure 1, the prompt provided to the LLM combines a predefined static instruction with a concretized dialogue template. The static instruction describes the task, output requirements, and language constraints. The concretized template is created by filling goal-conditioned slots with user-specific information sampled from behavioral records, including historical interaction items and their associated review texts. To strengthen the connection between dialogue content and product information, we incorporate real user reviews while instructing the model to summarize rather than reproduce them verbatim.

We use GPT-5o-nano to generate all dialogues in BanglaShop-CRS. The conversations follow the structure specified by the dialogue templates, including asking for recommendations, suggesting products, providing item details, and expressing acceptance or rejection. Leveraging the multilingual capabilities of the LLM, the generated utterances integrate product information from review texts in fluent Bangla, improving dialogue diversity and authenticity. Grounding generation in real reviews also preserves semantic consistency between the conversations and users' historical behaviors, which is important for reliable conversational recommendation evaluation [21, 22]. Appendix A1.1 Figure A1 shows examples of user profile and generated dialogues.

## 3.2 Dialogue Filtering

Due to the stochastic nature of LLM generation and the use of lengthy review texts, some generated dialogues may contain incomplete, noisy, or low-quality content. To ensure the reliability of BanglaShop-CRS, we apply a multi-step automatic filtering procedure before finalizing the dataset. First, we remove dialogues that are incomplete or truncated during generation, since partial conversations cannot represent the full recommendation process. We also discard outputs containing garbled, unreadable, or non-Unicode characters, which indicate generation failure. Dialogues with unfilled template slots are removed because they show that user-specific information was not properly substituted into the prompt. In addition, we filter out dialogues whose length falls outside the expected range of their corresponding template configuration, ensuring that each retained dialogue follows the intended multi-turn structure. To maintain linguistic consistency, we remove dialogues containing substantial non-Bangla content, as these violate the target language requirement. Finally, we discard dialogues that fail to incorporate user-specific review information, since such cases lack the behavioral grounding central to the user-centric design of BanglaShop-CRS.

# 4 Dataset Statistics

BanglaShop-CRS is constructed from the BanglishRev dataset [23], covering 130 products across 10 root categories. As shown in Table 2, the dataset consists of 27,178 multi-turn dialogues with 274,802 utterances and 3.6M tokens in total. On average, each dialogue contains approximately 10 utterances, consistent with our three-tier template design: Recommend Once (6--12 turns), Recommend Twice (8--14 turns), and Recommend Three Times (10--16 turns). A distinctive characteristic of BanglaShop-CRS is its user-centric design, in which each of the 273 users is associated with approximately 100 dialogue sessions (99.6 on average), enabling models to capture longitudinal preference patterns across multiple conversations. The dataset exhibits a natural domain imbalance: Health & Beauty, Groceries, and Home & Lifestyle contribute the largest share of dialogues and users, directly reflecting the interaction density distribution of the underlying BanglishRev data. Despite this imbalance, the average number of dialogues per user remains consistent across all 10 categories, confirming that the construction pipeline treats all users uniformly regardless of domain. BanglaShop-CRS is predominantly Bangla (92.16%), while retaining 7.75% code-mixed Bangla--English utterances, with only 0.07% Banglish and 0.01% English content (see Appendix A1.3 Table Table A2).

Table 2: Category-wise statistics of BanglaShop-CRS across 10 product domains. #4-Grams and #Unique 4-Grams reflect the lexical richness and semantic diversity of the generated dialogues.

| Root Category | #Dialogues | # Utter-ances | # Tokens | #4-Grams | #Unique 4-Grams | # Users |
|---|---|---|---|---|---|---|
| Electronic Devices | 1,331 | 13,374 | 180,140 | 176,147 | 142,041 | 14 |
| Groceries | 6,286 | 63,580 | 841,641 | 822,783 | 585,500 | 63 |
| Health & Beauty | 8,980 | 90,828 | 1,203,552 | 1,176,613 | 795,962 | 90 |
| Home & Lifestyle | 6,586 | 66,650 | 890,719 | 870,961 | 617,724 | 66 |
| Men's & Boys' Fashion | 997 | 10,088 | 129,555 | 126,564 | 101,073 | 10 |
| Mother & Baby | 899 | 9,087 | 120,035 | 117,338 | 92,671 | 9 |
| Sports & Outdoors | 200 | 2,000 | 25,921 | 25,321 | 21,596 | 2 |
| TV & Home Appliances | 199 | 1,999 | 26,577 | 25,980 | 22,564 | 2 |
| Watches, Bags & Jewellery | 900 | 9,085 | 118,054 | 115,354 | 91,562 | 9 |
| Women's & Girls' Fashion | 800 | 8,111 | 109,708 | 107,308 | 86,345 | 8 |
| **Total** | **27,178** | **274,802** | **3,645,902** | **3,564,369** | **2,557,038** | **273** |

### 4.1 Construction Cost.

The main cost of constructing BanglaShop-CRS comes from GPT-5-nano API calls during dialogue generation, with one call per dialogue and approximately 100,000 calls across preliminary generations. At $0.05 per 1M input tokens and $0.40 per 1M output tokens, an average prompt length of 1K tokens results in an estimated input cost of $5.00. Approximately 13.42M generated output tokens add $5.37, yielding a total estimated API cost of $10.37.

## 5 Evaluation

### 5.1 Evaluation on Conversational Recommendation

We conduct a series of experiments across all 10 product domains of BanglaShop-CRS to demonstrate its applicability to the conversational recommendation task and to highlight the importance of user-centric dialogues grounded in real interaction histories. Since generating fluent dialogue text is no longer a particularly challenging task for modern LLMs, our evaluation focuses primarily on the recommendation quality.

### 5.2 Evaluation Metrics.

We evaluate recommendation quality using Mean Reciprocal Rank (MRR), Recall@$K$, and NDCG@$K$ for $K \in \{5, 10, 50\}$. Let $r_i$ denote the rank of the target item for user $i$; if the target item is not retrieved, all scores are set to zero. MRR measures how early the correct item appears in the ranked recommendation list:

$$RR_i = \begin{cases} 1/r_i, & \text{if the target item is retrieved,} \\ 0, & \text{otherwise,} \end{cases}$$

$$\text{MRR} = \frac{1}{N} \sum_{i=1}^{N} RR_i.$$

Recall@$K$ measures whether the target item appears within the top $K$ recommendations:

$$\text{Recall@}K = \frac{1}{N} \sum_{i=1}^{N} \mathbf{1}(r_i \leq K).$$

NDCG@$K$ additionally rewards correct items that appear higher in the ranking:

$$\text{NDCG@}K = \frac{1}{N} \sum_{i=1}^{N} \frac{\mathbf{1}(r_i \leq K)}{\log_2(r_i + 1)}.$$

### 5.3 Baselines.

To verify the usability of BanglaShop-CRS, we evaluate four representative LLM-based baselines that support Bangla. Llama 3 (Llama-3-8B-Instruct) [24] and Qwen2.5 (Qwen2.5-7B-Instruct) [25] are strong open-source foundational models with multilingual capabilities. TigerLLM (TigerLLM-9B-it) [26] is a Bangla-specialized model built through continual pretraining on a Bangla textbook corpus followed by instruction tuning on a Bangla instruction dataset, using LLaMA 3.2 and Gemma-2 as backbone models. BanglaLLaMA (BanglaLLama-3.2-3b-bangla-alpaca-orca-instruct-v0.0.1) [27] is a causal language model fine-tuned on the Bangla-Alpaca-Orca dataset, based on LLaMA 3.2 3B. All models use deterministic greedy decoding (`do_sample=False`) without temperature scaling to ensure reproducibility and comparability across test users.

### 5.4 Input and Fine-Tuning Settings.

We evaluate each model under four settings: (1) *H.I Only*, using purchase history alone; (2) *H.I + Dial.*, using purchase history and dialogue context; (3) *Finetuned + H.I*, where the model is fine-tuned on all 27,178 BanglaShop-CRS dialogues and evaluated with purchase history; and (4) *Finetuned + H.I + Dial.*, where the fine-tuned model also receives dialogue context. We evaluate on 100 test users disjoint from the 273 users used for dataset construction and fine-tuning. For each user, $target_recommendation.item_name$ serves as the held-out target and is never shown during inference.

### 5.5 Recommendation Protocols.

We evaluate each setting under two protocols, following prior work on both free generation [11, 28] and candidate-based recommendation [29]. In the *catalog-constrained protocol*, the model receives the complete BanglaShop-CRS catalog and ranks 50 unique products from the full inventory. Unlike standard candidate-ranking settings, the catalog is not reduced to a small subset containing the target and a few distractors. Recommendations are then parsed, deduplicated, and checked against the catalog, with invalid items removed. If fewer than 50 valid products remain, up to two repair prompts are used to complete the list from the remaining catalog items. In the *open-vocabulary protocol*, no catalog is provided and the model freely generates ranked product names.

### 5.6 Product Matching and Failure Handling.

Following prior generative recommendation evaluations [4, 11], we use a sequential three-stage procedure to match generated recommendations to the held-out target. First, predictions and targets are normalized by lowercasing, replacing `&" with `and," removing punctuation outside Latin and Bangla characters and digits, and collapsing repeated whitespace; identical normalized strings are treated as exact matches. Second, if exact matching fails, we apply bidirectional substring matching, where either normalized string must be fully contained within the other. Third, we compute RapidFuzz `token_set_ratio`, which reduces sensitivity to word order and repeated terms, and treat scores of at least 85 as matches. Predictions are examined in ranked order, and the first matched rank is used to compute MRR, Recall@$K$, and NDCG@$K$. If no recommendation matches, the example remains in the denominator and contributes zero to all metrics. Missing targets are excluded from both protocols, while targets absent from the catalog are additionally excluded from catalog-constrained evaluation because retrieval is impossible by construction. The complete algorithm and prompts are provided in Appendix A1.4.

## 6 Results and Discussions

Table 3 presents baseline and fine-tuned performance under the catalog-based protocol. Adding dialogue context improves every non-fine-tuned model, increasing MRR from 0.0609 to 0.0690 for Llama 3, 0.0589 to 0.0771 for Qwen, 0.0429 to 0.0594 for TigerLLM, and 0.0372 to 0.0696 for BanglaLLaMA. The largest gain is for BanglaLLaMA (approximately 87%), showing that dialogue provides useful preference signals beyond purchase history. Fine-tuning further improves performance, with the Finetuned + H.I + Dial. setting achieving the highest MRR for all models: 0.1081 for Llama 3, 0.1174 for Qwen, 0.1107 for TigerLLM, and 0.0796 for BanglaLLaMA, corresponding to gains of approximately 57%, 52%, 86%, and 14%, respectively. Qwen achieves the highest MRR, R@5, N@5, and N@10, while TigerLLM obtains the highest R@10 and N@50 and, under Finetuned + H.I, the highest R@50. TigerLLM's strong broader-cutoff performance suggests that its Bangla-specific pretraining may support retrieval of a wider range of culturally and linguistically relevant products.

Open-vocabulary results are reported in Appendix A2 and Table A5. Under the Finetuned + H.I + Dial. setting, catalog-based MRR is higher for Qwen, TigerLLM, and BanglaLLaMA by approximately 67%, 22%, and 99%, respectively, while Llama 3 performs similarly across both protocols. This advantage arises because catalog-based evaluation ranks predefined products, whereas open-vocabulary evaluation requires generating product references that can be matched to the catalog.

### 6.1 Human Evaluation

We conduct a human evaluation to assess the quality and reliability of BanglaShop-CRS at both the utterance and conversation levels. Because human judgments may be influenced by individual preferences and contextual familiarity, we recruit five volunteer annotators who are native Bangla speakers. Before annotation, all annotators are informed of the research objectives and briefed on the evaluation criteria and rating standards.

### 6.2 Utterance-Level Evaluation.

We randomly sample 100 dialogues from BanglaShop-CRS and shuffle their order to reduce ordering bias. Each annotator evaluates all 960 utterances across four dimensions: (1)

Table 3: Recommendation performance of the baseline and fine-tuned models under the catalog-based evaluation protocol on BanglaShop-CRS. H.I denotes purchase history, while Dial. denotes dialogue context from the dataset. Orange rows indicate non-fine-tuned models, whereas blue rows indicate fine-tuned models.

| Model | Setting | MRR | R@5 | R@10 | R@50 | N@5 | N@10 | N@50 |
|---|---|---|---|---|---|---|---|---|
| Llama 3 | H.I only | 0.0609 | 0.08 | 0.12 | 0.30 | 0.0547 | 0.0665 | 0.1086 |
| | H.I + Dial. | 0.0690 | 0.11 | 0.16 | 0.31 | 0.0680 | 0.0835 | 0.1193 |
| | Finetuned + H.I | 0.0969 | 0.16 | 0.25 | 0.40 | 0.0990 | 0.1265 | 0.1602 |
| | Finetuned + H.I + Dial. | 0.1081 | 0.17 | 0.24 | 0.37 | 0.1107 | 0.1328 | 0.1643 |
| Qwen | H.I only. | 0.0589 | 0.08 | 0.14 | 0.20 | 0.0552 | 0.0751 | 0.0898 |
| | H.I+Dial. | 0.0771 | 0.11 | 0.19 | 0.29 | 0.0718 | 0.0985 | 0.1232 |
| | Finetuned+H.I | 0.1048 | 0.17 | 0.22 | 0.34 | 0.1105 | 0.1264 | 0.1555 |
| | Finetuned+H.I + Dial. | 0.1174 | 0.18 | 0.24 | 0.36 | 0.1205 | 0.1401 | 0.1701 |
| TigerLLM | H.I only | 0.0429 | 0.06 | 0.11 | 0.18 | 0.0402 | 0.0560 | 0.0715 |
| | H.I+Dial. | 0.0594 | 0.08 | 0.14 | 0.24 | 0.0552 | 0.0739 | 0.0973 |
| | Finetuned +H.I | 0.0941 | 0.13 | 0.24 | 0.46 | 0.0821 | 0.1180 | 0.1702 |
| | Finetuned +H.I + Dial. | 0.1107 | 0.15 | 0.25 | 0.43 | 0.1031 | 0.1351 | 0.1770 |
| Bangla LLaMA | H.I Only | 0.0372 | 0.04 | 0.08 | 0.18 | 0.0299 | 0.0503 | 0.0649 |
| | H.I + Dial. | 0.0696 | 0.12 | 0.19 | 0.26 | 0.05 | 0.062 | 0.1031 |
| | Finetuned+ H.I . | 0.0472 | 0.07 | 0.12 | 0.20 | 0.0442 | 0.0603 | 0.0794 |
| | Finetuned+ H.I +Dial | 0.0796 | 0.14 | 0.21 | 0.33 | 0.0817 | 0.1042 | 0.1331 |

**Fluency**, measuring grammatical correctness and readability in Bangla; (2) **Informativeness**, measuring whether responses are meaningful rather than generic or repetitive; (3) **Logicality**, assessing logical consistency and common-sense reasoning; and (4) **Coherence**, assessing consistency with preceding dialogue turns. Each utterance is rated on a 0--5 scale for every dimension. The detailed rating scheme is provided in Appendix A1.5.

Table 4: Utterance-level human evaluation scores and inter-annotator agreement across four quality dimensions. Significance stars indicate the chi-square significance test of Kendall's $W$: $^{***}p < 0.001$, $^{**}p < 0.01$, $^{*}p < 0.05$.

| Metric | Avg. Score | Kendall's $W$ | $\chi^2$ |
|---|---|---|---|
| Fluency | 3.598 | $0.843^{***}$ | 417.358 |
| Informativeness | 3.076 | $0.832^{***}$ | 411.800 |
| Logicality | 2.901 | $0.407^{***}$ | 201.679 |
| Coherence | 2.882 | $0.809^{***}$ | 400.209 |

## 6.3 Results and Analysis.

We use Kendall's coefficient of concordance ($W$) to measure inter-annotator agreement among the five annotators across four utterance-level dimensions: fluency, informativeness, logicality, and coherence. We test the significance of $W$ using the standard chi-square procedure. As shown in Table 4, all four dimensions achieve statistically significant agreement at $p < 0.001$, indicating consistent annotator judgments. Agreement is particularly strong for fluency ($W = 0.843$, $\chi^2 = 417.358$), informativeness ($W = 0.832$, $\chi^2 = 411.800$), and coherence ($W = 0.809$, $\chi^2 = 400.209$). Logicality shows lower but still significant agreement ($W = 0.407$, $\chi^2 = 201.679$), which is expected because logical consistency requires more subjective interpretation than fluency or surface-level coherence. Overall, these agreement values suggest that the human evaluation results are reliable and not due to chance.

Table 4 shows that BanglaShop-CRS achieves strong overall quality. Fluency receives the highest score (3.598), followed by informativeness (3.076), indicating generally natural and useful product-related dialogue. Logicality (2.901) and coherence (2.882) are slightly lower, reflecting occasional weak alignment or topic shifts. Overall, grounding the dialogues in user histories and reviews supports fluent, informative, and user-centered Bangla recommendations.

## 6.4 Conversation-Level Evaluation.

We evaluate overall dialogue quality using three labels: *poor*, *normal*, and *good*. The annotation guidelines are provided in Appendix A1.6. The same 100 dialogues used for utterance-level evaluation are assessed by five annotators, producing 500 annotations. Kendall's $W$ indicates moderate and statistically significant agreement ($W = 0.463$, $\chi^2(99) = 228.945$, $p < 0.001$). As shown in Figure 2, most dialogues are rated *normal* or *good*, with few receiving a *poor* rating. These findings suggest that the template-guided, review-grounded pipeline generally produces coherent and contextually appropriate Bangla recommendation dialogues.

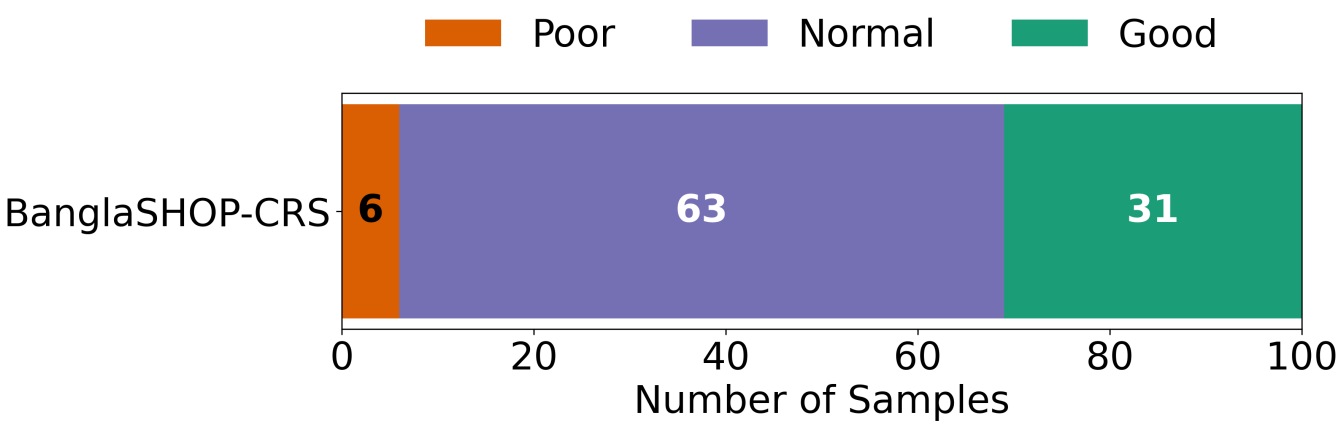

Figure 2: Conversation-level quality distribution based on majority voting from five native Bangla-speaking annotators. The stacked bar shows the proportion of dialogue samples rated as *poor*, *normal*, and *good* across the 100 sampled dialogues from BanglaShop-CRS.

## 6.5 Template Diversity Analysis.

We quantify the structural diversity of the 100 dialogue templates using recommendation count, dialogue length, and ordered speaker, goal, and subgoal sequences (See Appendix A2.1 Table A6). The templates are nearly evenly distributed across one-, two-, and three-recommendation settings (34, 33, and 33 templates, respectively) and span 6--16 turns. We identify 84 unique goal sequences, 100 unique subgoal sequences, and 45 unique speaker sequences. Normalized Shannon entropy is high for goal (0.988), subgoal (1.000), and speaker (0.935) sequences, indicating broad structural coverage. Mean normalized edit distance is highest for subgoal sequences (0.722), followed by goal sequences (0.444) and speaker sequences (0.266), reflecting greater variation in conversational functions than in the alternating user--agent structure.(See Appendix A2.1 for details).

## 6.6 Factual Grounding Evaluation.

To directly assess factual grounding, three native Bangla-speaking annotators evaluated 200 sampled dialogue claims against the corresponding product, user rating, and source review using a 0--4 scale, where 0 indicates unsupported or contradictory content and 4 indicates full support (See appnedix A3 for annotation guideline). The claims were also evaluated against shuffled user records as a control. The correct-user condition achieved an average score of 2.87 (Mostly supported), compared with 1.35(Weekly related) for the shuffled-user condition, indicating stronger alignment with the intended user histories. Inter-annotator agreement was substantial ($\kappa = 0.54$). We additionally used GPT-5.1 as an independent judge (See appnedix A3.3 for the prompting strategy)), which produced an average grounding score of 3.08. The similar human and LLM scores provide further evidence that the generated dialogues are generally grounded in the source reviews and user preferences.

# 7 Conclusion

We introduced BanglaShop-CRS, a large-scale user-centric Bangla conversational recommendation dataset grounded in real e-commerce histories, ratings, and reviews. The dataset contains 27,178 multi-turn dialogues across 10 product domains and is generated through a template-guided LLM pipeline. Experiments under catalog-constrained and open-vocabulary protocols show that dialogue context improves recommendation quality and that fine-tuning on BanglaShop-CRS yields further gains across Bangla-capable LLMs. Human evaluation confirms the fluency, informativeness, logicality, and coherence of the dialogues, while factual-grounding evaluation shows stronger alignment with correct user records than shuffled controls. Overall, BanglaShop-CRS provides a scalable benchmark for advancing conversational recommendation in Bangla.

# Limitations

BanglaShop-CRS has several limitations. Dialogue quality depends on the selected prompts and manually designed goals and subgoals. Although the 100 templates provide broad structural coverage, recurring dialogue trajectories may still encourage template-specific learning and limit scalability. Future work should explore automatic goal induction, prompt optimization, and less constrained dialogue generation. The dataset also inherits noise and bias from the underlying ratings and reviews. Fixed rating thresholds may oversimplify mixed preferences, while reviews can contain incomplete, exaggerated, or subjective claims. Moreover, evaluating against a single held-out target may treat other relevant products as false negatives. Multiple or graded relevance labels would provide a more complete evaluation.

The final dataset is predominantly Bangla, with relatively limited Banglish, English, and code-mixed content. It may therefore underrepresent transliterated Bangla, regional dialects, informal spelling, and naturally occurring language switching. The dataset is also limited to 273 users and 130 products across 10 domains. This controlled scale underrepresents long-tail products, sparse users, cold-start settings, and the retrieval difficulty of large commercial catalogs. The results should therefore be interpreted as evidence within a controlled Bangla CRS benchmark rather than deployment-level generalizability. Our direct grounding analysis verifies whether product mentions align with users' recorded likes, dislikes, and recommendation candidates, but it does not validate every claim against the original review text. A larger claim-level evaluation labeling statements as supported, contradicted, or unverifiable would provide stronger factual-grounding evidence. Similarly, catalog-based evaluation ensures valid product outputs but uses a relatively small candidate set, whereas open-vocabulary evaluation remains sensitive to lexical variation and fuzzy-matching decisions.

Finally, generation with GPT-5-nano may introduce linguistic, cultural, commercial, or demographic biases [30, 31]. Performance may also be weaker for Bangla dialects, Banglish, informal reviews, and culturally specific shopping contexts [32]. Some dialogues may therefore contain unsupported elaborations, weak contextual alignment, or biased recommendation patterns. Future work should strengthen factuality checking, bias detection, and evaluation with larger and more diverse native-speaker samples.

## Ethics Considerations

BanglaShop-CRS is constructed from the BanglishRev dataset [23], which contains product reviews collected from e-commerce platforms for Bengali-speaking users. BanglishRev is publicly available for research and does not include private information such as real names, phone numbers, or physical addresses. In our pipeline, we use only anonymized reviewer identifiers, ensuring that individual consumers cannot be identified from BanglaShop-CRS. The dialogues in BanglaShop-CRS are synthetically generated by an LLM and do not reproduce review content verbatim. They are grounded in user interaction patterns and review signals rather than direct personal disclosures. The five native Bangla-speaking annotators participated voluntarily, were informed of the research objectives, and were briefed on the evaluation criteria before annotation. No sensitive or personally identifiable information was shown during evaluation. BanglaShop-CRS is intended for research purposes only. It may not be used for commercial applications or redistributed without permission. Dataset access will follow a structured application process requiring prospective users to provide their name, affiliation, research area, and intended use case. Access will be granted after review and approval by the authors.

## Acknowledgements

The authors are grateful for National Science Foundation Award #2242829 (Science of Online Corpora, Knowledge, and Stories), foundational support from MassMutual, and an anonymous philanthropic gift.

## References

[1] J. Shi, R. Jain, H. Doh, R. Suzuki, and K. Ramani. An hci-centric survey and taxonomy of human-generative-ai interactions. *ArXiv*, abs/2310.07127, 2023.

[2] M. Lee, P. Liang, and Q. Yang. Coauthor: Designing a human-ai collaborative writing dataset for exploring language model capabilities. *Proceedings of the 2022 CHI Conference on Human Factors in Computing Systems*, 2022.

[3] K. Zhou, W. X. Zhao, S. Bian, Y. Zhou, J. rong Wen, and J. Yu. Improving conversational recommender systems via knowledge graph based semantic fusion. *Proceedings of the 26th ACM SIGKDD International Conference on Knowledge Discovery & Data Mining*, 2020.

[4] Z. He, Z. Xie, R. Jha, H. Steck, D. Liang, Y. Feng, B. P. Majumder, N. Kallus, and J. McAuley. Large language models as zero-shot conversational recommenders. *Proceedings of the 32nd ACM International Conference on Information and Knowledge Management*, 2023.

[5] K. Zhou, X. Wang, Y. Zhou, C. Shang, Y. Cheng, W. X. Zhao, Y. Li, and J. rong Wen. Crslab: An open-source toolkit for building conversational recommender system. In *Annual Meeting of the Association for Computational Linguistics*, 2021.

[6] Y. Deldjoo, Z. He, J. McAuley, A. Korikov, S. Sanner, A. Ramisa, R. Vidal, M. Sathiamoorthy, A. Kasirzadeh, and S. Milano. A review of modern recommender systems using generative models (gen-recsys). *Proceedings of the 30th ACM SIGKDD Conference on Knowledge Discovery and Data Mining*, 2024.

[7] S. Yun and Y. kyung Lim. User experience with llm-powered conversational recommendation systems: A case of music recommendation. *Proceedings of the 2025 CHI Conference on Human Factors in Computing Systems*, 2025.

[8] R. Li, S. E. Kahou, H. Schulz, V. Michalski, L. Charlin, and C. J. Pal. Towards deep conversational recommendations. *ArXiv*, abs/1812.07617, 2018.

[9] K. Zhou, Y. Zhou, W. X. Zhao, X. Wang, and J. rong Wen. Towards topic-guided conversational recommender system. *ArXiv*, abs/2010.04125, 2020.

[10] Z. Liu, H. Wang, Z.-Y. Niu, H. Wu, W. Che, and T. Liu. Towards conversational recommendation over multi-type dialogs. In *Annual Meeting of the Association for Computational Linguistics*, 2020.

[11] T. Liang, C. Jin, L. Wang, W. Fan, C. Xia, K. Chen, and Y. Yin. Llm-redial: A large-scale dataset for conversational recommender systems created from user behaviors with llms. In *Annual Meeting of the Association for Computational Linguistics*, 2024.

[12] J. N. Pava, C. Meinhardt, H. Badi Uz Zaman, T. Friedman, S. T. Truong, D. Zhang, E. Cryst, V. Marivate, and S. Koyejo. Mind the (language) gap: Mapping the challenges of llm development in low-resource language contexts. White paper, Stanford Institute for Human-Centered Artificial Intelligence, The Asia Foundation, and University of Pretoria, Apr. 2025.

[13] Central Intelligence Agency. Field Listing -- Ethnic Groups. https://www.cia.gov/the-world-factbook/field/ethnic-groups/, 2025. Retrieved January 18, 2025.

[14] Wikipedia. Bengalis --- Wikipedia, The Free Encyclopedia. https://en.wikipedia.org/w/index.php?title=Bengalis&oldid=1268353351, 2025. Retrieved January 18, 2025.

[15] M. Mostafa and M. Jamila. From english to banglish: Loanwords as opportunities and barriers? *English Today*, 28:26 -- 31, 2012.

[16] T. T. Prama, C. M. Danforth, and P. S. Dodds. Banglamath : A bangla benchmark dataset for testing llm mathematical reasoning at grades 6, 7, and 8. *ArXiv*, abs/2510.12836, 2025.

[17] C. Alexandris. Genai and socially responsible ai in natural language processing applications: A linguistic perspective. In *AAAI Spring Symposia*, 2024.

[18] Y. J. Choi, M. Lee, and S. Lee. Toward a multilingual conversational agent: Challenges and expectations of code-mixing multilingual users. *Proceedings of the 2023 CHI Conference on Human Factors in Computing Systems*, 2023.

[19] F. Kruk, S. Herath, and P. R. Choudhury. Banglassist: A bengali-english generative ai chatbot for code-switching and dialect-handling in customer service. *Proceedings of the Extended Abstracts of the CHI Conference on Human Factors in Computing Systems*, 2025.

[20] X. V. Yu, A. Asai, T. Chatterjee, J. Hu, and E. Choi. Beyond counting datasets: A survey of multilingual dataset construction and necessary resources. *ArXiv*, abs/2211.15649, 2022.

[21] X. Wang, K. Zhou, J. rong Wen, and W. X. Zhao. Towards unified conversational recommender systems via knowledge-enhanced prompt learning. *Proceedings of the 28th ACM SIGKDD Conference on Knowledge Discovery and Data Mining*, 2022.

[22] I. Kostric, K. Balog, and U. Gadiraju. Should we tailor the talk? understanding the impact of conversational styles on preference elicitation in conversational recommender systems. *Proceedings of the 33rd ACM Conference on User Modeling, Adaptation and Personalization*, 2025.

[23] M. N. Shamael, S. Nawshin, S. Shatabda, and S. Islam. Banglishrev: A large-scale bangla-english and code-mixed dataset of product reviews in e-commerce. *ArXiv*, abs/2412.13161, 2024.

[24] K. C. Huang, K. Lakhotia, K. Huang, L. Chen, L. Garg, A. Lavender, L. Silva, L. Bell, L. Zhang, L. Guo, et al. The llama 3 herd of models. *preprint*, 2024.

[25] S. Bai, K. qin Chen, X. Liu, J. Wang, W. Ge, S. Song, K. Dang, P. Wang, S. Wang, J. Tang, et al. Qwen2.5-vl

technical report. *ArXiv*, abs/2502.13923, 2025.

[26] N. Raihan and M. Zampieri. Tigerllm - a family of bangla large language models. *ArXiv*, abs/2503.10995, 2025.

[27] A. K. Zehady, S. R. Dipta, N. Islam, S. M. A. A. Mamun, and S. Karmaker. Banglallama: Llama for bangla language. *Proceedings of the Second Workshop on Language Models for Low-Resource Languages (LoResLM 2026)*, 2024.

[28] J. Liu, C. Liu, R. Lv, K. Zhou, and Y. B. Zhang. Is chatgpt a good recommender? a preliminary study. *ArXiv*, abs/2304.10149, 2023.

[29] J. Lin, X. Dai, Y. Xi, W. Liu, B. Chen, X. Li, C. Zhu, H. Guo, Y. Yu, R. Tang, and W. Zhang. How can recommender systems benefit from large language models: A survey. *ACM Transactions on Information Systems*, 43:1 -- 47, 2023.

[30] T. T. Prama and M. S. Islam. Evaluating credibility and political bias in llms for news outlets in bangladesh. In *Annual Meeting of the Association for Computational Linguistics*, 2025.

[31] T. T. Prama, J. W. Zimmerman, C. M. Danforth, and P. S. Dodds. Us-vs-them bias in large language models. *ArXiv*, abs/2512.13699, 2025.

[32] T. T. Prama, C. M. Danforth, and P. S. Dodds. Llms for low-resource dialect translation using context-aware prompting: A case study on sylheti. *ArXiv*, abs/2511.21761, 2025.

# A1 Appendices

## A1.1 Example of Dialogue

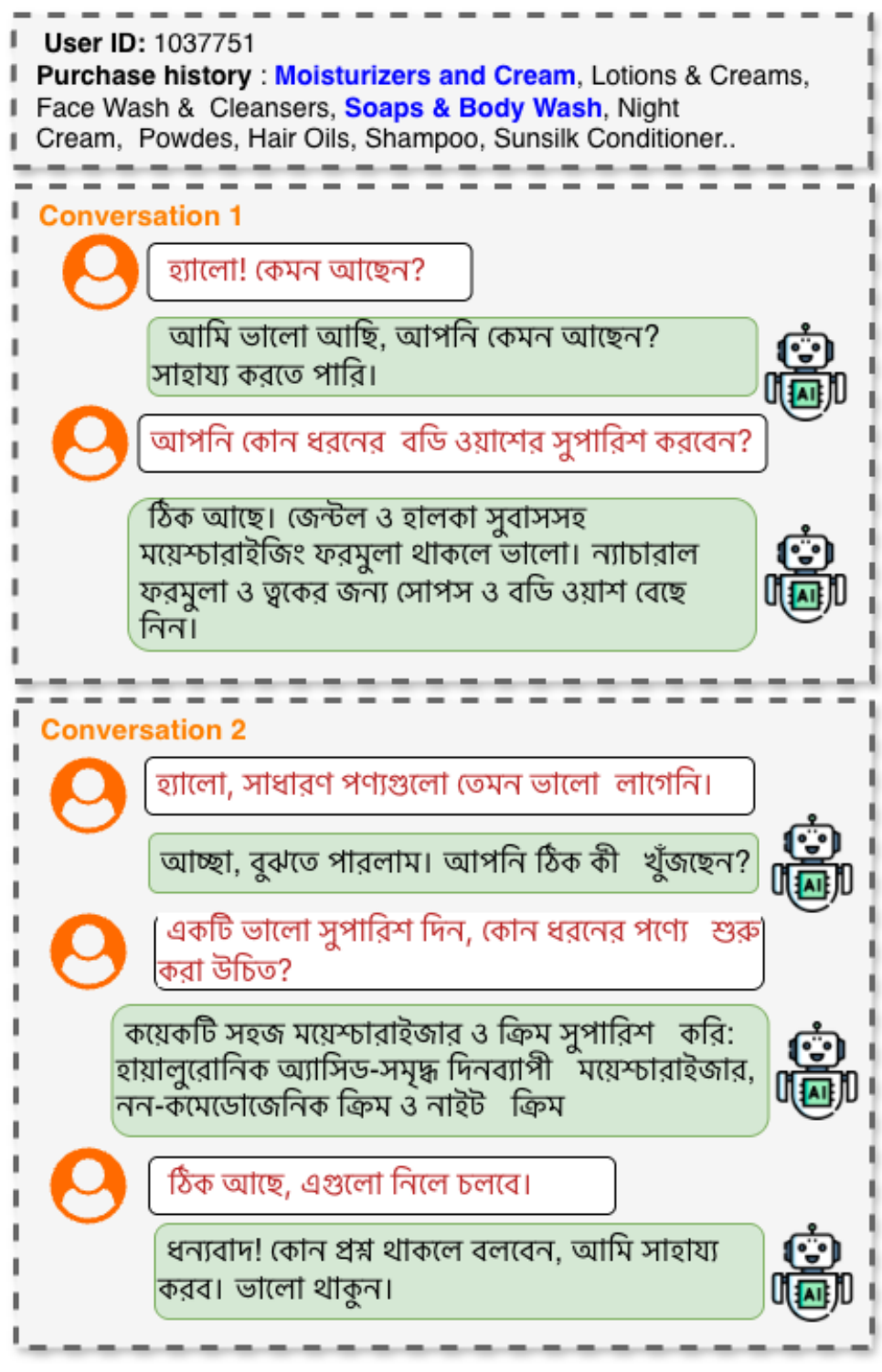

Figure A1: Example user profile and LLM-generated dialogues in BanglaShop-CRS. A single user may be associated with multiple conversations, and the items mentioned in the dialogues are consistent with the user's historical interactions.

## A1.2 Goals and Subgoals

Table A1 shows 8 primary goals and 25 subgoals with the related descriptions.

## A1.3 Language composition

## A1.4 Prompting Strategy

Tables A3 and A4 present the prompts used for conversational recommendation. We design two prompting settings: one using only the user's purchase history, and another using both purchase history and dialogue context. Since the user interactions in BanglaShop-CRS are generated from Bangla and Banglish reviews, the prompts are written to preserve Bangla conversational context while requiring the model to produce English product recommendations.

## A1.5 Annotation Guidelines

We evaluate each dialogue using a 0--5 scale across four dimensions: fluency, informativeness, logical consistency, and coherence.

Table A1: Primary dialogue goals and 25 representative sub-goals and utterances descriptions in BanglaShop-CRS templates.

| Primary Goal | Sub-Goal | Utterances Description |
|---|---|---|
| Greeting | Simple greeting | Starts with a basic greeting such as hello, hi, or salam. |
| | Greeting with preference cue | Opens the dialogue using the user's preference signal. |
| | Greeting with contextual information | Starts with context from the user's history or need. |
| Ask | Ask for recommendation | The user asks for product recommendations. |
| | Ask for more information or clarification | The user requests additional information. |
| | Ask for alternatives or comparison | The user asks for other options or comparisons. |
| | Ask with constraints or preferences | The user asks with specific requirements. |
| Respond | Direct response | The system gives a direct answer. |
| | Informative elaboration | The system provides additional information. |
| | Clarification-oriented response | The system clarifies the user's need. |
| | Constraint-aware response | The system responds based on user constraints. |
| | Comparative response | The system compares available options. |
| Recommend | Single-item recommendation | The system recommends one item. |
| | Multiple-item recommendation | The system recommends multiple items. |
| | Justification-supported recommendation | The system explains why an item is recommended. |
| | Personalized recommendation | The system recommends based on user history or preference. |
| Feedback | Acceptance or rejection | The user accepts or rejects a recommendation. |
| | Refinement request after dissatisfaction | The user asks for a better recommendation. |
| | Dynamic preference update | The user updates preferences during the conversation. |
| Chit-Chat | Casual social interaction | Adds natural conversational interaction. |
| Reason | Direct answering | The user gives a direct reason. |
| | Informative elaboration | The user explains the reason in more detail. |
| | Clarification-seeking response | The user asks for clarification while reasoning. |
| | Constraint-aware response | The user explains rejection based on constraints. |
| | Comparative response | The user compares the recommendation with alternatives. |

**Fluency.** Fluency measures whether the dialogue is grammatically correct, natural, and easy to understand.

- **0 (very poor):** The dialogue contains severe grammatical errors, spelling mistakes, vocabulary problems, or incoherent expressions that make it very difficult or impossible to understand.
- **1 (poor):** The dialogue contains frequent grammar errors, spelling mistakes, or awkward word choices. The meaning can sometimes be understood, but comprehension requires significant effort.
- **2 (fair):** The dialogue contains several noticeable grammar, spelling, or vocabulary issues. However, the overall meaning is mostly understandable.
- **3 (normal):** The dialogue is generally understandable but may contain some grammar errors, spelling mistakes, awkward phrasing, or minor fluency issues.
- **4 (good):** The dialogue is mostly fluent and clear, with only minor grammatical, spelling, or vocabulary issues that do not affect comprehension.
- **5 (excellent):** The dialogue is fully fluent, natural, and coherent, with no noticeable grammar errors, spelling mistakes, or vocabulary problems.

Table A2: Language composition of utterances in BanglaShop-CRS.

| Language | Count | Percentage (%) |
|---|---|---|
| Bangla | 253,244 | 92.16 |
| Code-mixed | 21,320 | 7.75 |
| Banglish | 215 | 0.07 |
| English | 23 | 0.01 |

Table A3: Prompts used for product recommendation generation using user historical interactions as context only.

| H.I Prompt in Bangla | H.I Prompt translated in English Prompt |
|---|---|
| ধরুন আপনি একজন ই-কমার্স রিকমেন্ডার সিস্টেম।<br><br>আমি আপনাকে একজন test user's previous interaction with product history দেব।<br>এই test user's history এবং conversation context দেখে ঠিক ৫০টি product recommendation দিন।<br><br>**নিয়ম:**<br>• সব পণ্যের নাম অবশ্যই English-এ লিখবেন।<br>• শুধু product name বা product category লিখবেন।<br>• প্রতিটি product আলাদা লাইনে লিখবেন।<br>• প্রতিটি লাইন অবশ্যই number + dot + space দিয়ে শুরু হবে।<br>• Format must be exactly:<br>1. Product Name<br>2. Product Name<br>3. Product Name<br>…<br>50. Product Name<br>• এক লাইনে একাধিক product লিখবেন না।<br>• কোনো Bangla product name লিখবেন না।<br>• কোনো ব্যাখ্যা লিখবেন না।<br>• কোনো অতিরিক্ত বাক্য লিখবেন না।<br>• User: বা Agent: লিখবেন না।<br>• N/A লিখবেন না।<br>• product repeat করবেন না।<br>• ঠিক ৫০টি item দিন।<br><br>**Test User Interaction History:**<br>`{history_text}`<br><br>এখন শুধু ১ থেকে ৫০ পর্যন্ত numbered English product recommendations দিন: | Assume that you are an e-commerce recommender system.<br><br>I will provide a test user's previous interaction with product history. Based on this test user's history and conversation context, recommend exactly 50 products.<br><br>**Rules:**<br>• All product names must be written in English.<br>• Write only product names or product categories.<br>• Write each product on a separate line.<br>• Each line must start with number + dot + space.<br>• Format must be exactly:<br>1. Product Name<br>2. Product Name<br>3. Product Name<br>…<br>50. Product Name<br>• Do not put multiple products on one line.<br>• Do not write Bangla product names.<br>• Do not write explanations.<br>• Do not write extra sentences.<br>• Do not write User: or Agent:.<br>• Do not write N/A.<br>• Do not repeat products.<br>• Provide exactly 50 items.<br><br>**Test User Interaction History:**<br>`{history_text}`<br><br>Now return only numbered English product recommendations from 1 to 50. |

**Informativeness.** Informativeness measures whether the dialogue provides useful, specific, and decision-supporting information.

- **0 (very poor):** The dialogue provides almost no useful information. It mainly consists of vague, repeated, generic, or safe responses and does not help the user understand the product or recommendation.
- **1 (poor):** The dialogue provides very limited information. It gives a basic response but lacks useful details, explanations,

Table A4: Prompts used for product recommendation generation using both purchase history and dialogue sample ( H.I + Dial.).

| **H.I + Dial. Prompt in Bangla** | **H.I + Dial. Prompt translated in English** |
|---|---|
| ধরুন আপনি একজন ই-কমার্স রিকমেন্ডার সিস্টেম।<br><br>আমি আপনাকে একজন test user's previous purchase history এবং একটি Bangla conversation context দেব।<br>এই test user's history এবং conversation context দেখে ঠিক ৫০টি product recommendation দিন।<br><br>**নিয়ম:**<br>• সব পণ্যের নাম অবশ্যই English-এ লিখবেন।<br>• শুধু product name বা product category লিখবেন।<br>• প্রতিটি product আলাদা লাইনে লিখবেন।<br>• প্রতিটি লাইন অবশ্যই number + dot + space দিয়ে শুরু হবে।<br>• Format must be exactly:<br>1. Product Name<br>2. Product Name<br>3. Product Name<br>…<br>50. Product Name<br>• এক লাইনে একাধিক product লিখবেন না।<br>• কোনো Bangla product name লিখবেন না।<br>• কোনো ব্যাখ্যা লিখবেন না।<br>• কোনো অতিরিক্ত বাক্য লিখবেন না।<br>• User: বা Agent: লিখবেন না।<br>• N/A লিখবেন না।<br>• product repeat করবেন না।<br>• ঠিক ৫০টি item দিন।<br><br>**Test User Purchase History:**<br>`{history_text}`<br><br>**Bangla Conversation Context:**<br>`{User_Conversation}`<br><br>এখন শুধু ১ থেকে ৫০ পর্যন্ত numbered English product recommendations দিন: | Assume that you are an e-commerce recommender system.<br><br>I will provide a test user's previous purchase history and a Bangla conversation context. Based on this test user's history and conversation context, recommend exactly 50 products.<br><br>**Rules:**<br>• All product names must be written in English.<br>• Write only product names or product categories.<br>• Write each product on a separate line.<br>• Each line must start with number + dot + space.<br>• Format must be exactly:<br>1. Product Name<br>2. Product Name<br>3. Product Name<br>…<br>50. Product Name<br>• Do not put multiple products on one line.<br>• Do not write Bangla product names.<br>• Do not write explanations.<br>• Do not write extra sentences.<br>• Do not write User: or Agent:.<br>• Do not write N/A.<br>• Do not repeat products.<br>• Provide exactly 50 items.<br><br>**Test User Purchase History:**<br>`{history_text}`<br><br>**Bangla Conversation Context:**<br>`{User_Conversation}`<br><br>Now return only numbered English product recommendations from 1 to 50. |

product features, or relevant context needed for user understanding.

- **2 (fair):** The dialogue provides some relevant information, but the response remains shallow or incomplete. The user may still need additional details to make a decision.
- **3 (normal):** The dialogue provides generally useful information and answers the user's main request, but the explanation lacks depth, specificity, or additional supporting details.
- **4 (good):** The dialogue provides clear and relevant information with useful product details, reasons, features, or comparisons. It mostly supports user understanding and decision-making.
- **5 (excellent):** The dialogue provides rich, detailed, and highly relevant information. It fully answers the user's query, includes useful product-specific details, and offers additional context that supports an informed decision.

**Logical Consistency.** Logical consistency measures whether the dialogue responses and recommendations are reasonable, relevant, and contextually appropriate.

- **0 (very poor):** The dialogue contains severe logical errors, unrelated responses, clear contradictions, or completely irrelevant suggestions that make the conversation difficult to follow.
- **1 (poor):** The dialogue has frequent logical issues. Many responses or recommendations are weakly related to the user's query or context, though some parts remain understandable.
- **2 (fair):** The dialogue has some logical inconsistencies or off-topic shifts, but the main conversation can still be followed.
- **3 (normal):** The dialogue is mostly logical and understandable, but some responses or suggestions may be only loosely connected to the user's request.
- **4 (good):** The dialogue is logically coherent, with responses and recommendations that are relevant and reasonable in most parts of the conversation.
- **5 (excellent):** The dialogue is fully logical, coherent, and contextually appropriate. All responses and recommendations are clearly connected to the user's needs and previous turns.

**Coherence.** Coherence measures whether the dialogue maintains contextual continuity and smooth transitions across turns.

- **0 (very poor):** The dialogue is highly incoherent, with no clear contextual connection between turns. Responses or suggestions appear random, disconnected, or impossible to follow.
- **1 (poor):** The dialogue has major coherence problems. Some turns are understandable, but responses often shift topics abruptly or fail to connect with previous context.
- **2 (fair):** The dialogue shows partial coherence, but there are noticeable ruptures, weak transitions, or insufficient links between user requests and agent responses.
- **3 (normal):** The dialogue is generally coherent and understandable, but it may contain occasional topic shifts, weak contextual links, or slightly abrupt transitions.
- **4 (good):** The dialogue is coherent, with mostly clear connections between responses, suggestions, and prior context. Transitions are smooth in most parts.
- **5 (excellent):** The dialogue is highly coherent, with strong contextual continuity, clear logical links between turns, and smooth transitions throughout the conversation.

### A1.6 Conversation-level Quality.

Conversation-level quality evaluates the overall dialogue as a complete interaction, considering whether the conversation is natural, informative, coherent, and useful for recommendation.

- **0 (poor):** The conversation is difficult to follow, contains major topic shifts or irrelevant recommendations, and does not adequately address the user's needs.
- **1 (normal):** The conversation is generally understandable and partially relevant, but it may contain some weak transitions, limited information, or recommendations that are only moderately aligned with the user's request.
- **2 (good):** The conversation is coherent, relevant, and useful. The agent responds appropriately to the user's needs, provides meaningful recommendations, and maintains a smooth conversational flow.

Table A5: Recommendation performance of the baseline and fine-tuned models under the open-vocabulary evaluation protocol on BanglaShop-CRS. H.I denotes purchase history, while Dial. denotes dialogue context from the dataset. Orange rows indicate non-fine-tuned models, whereas blue rows indicate fine-tuned models.

| Model | Setting | MRR | R@5 | R@10 | R@50 | N@5 | N@10 | N@50 |
|---|---|---|---|---|---|---|---|---|
| Llama 3 | H.I Only | 0.0298 | 0.03 | 0.07 | 0.23 | 0.0193 | 0.0319 | 0.0694 |
| | H.I + Dial. | 0.0531 | 0.09 | 0.17 | 0.25 | 0.0516 | 0.0764 | 0.0950 |
| | Finetuned + H.I | 0.1099 | 0.15 | 0.26 | 0.38 | 0.1040 | 0.1391 | 0.1670 |
| | Finetuned + H.I + Dial. | 0.1128 | 0.13 | 0.27 | 0.37 | 0.1069 | 0.1447 | 0.1708 |
| Qwen | H.I Only | 0.0325 | 0.04 | 0.07 | 0.14 | 0.0293 | 0.0385 | 0.0538 |
| | H.I + Dial. | 0.0524 | 0.05 | 0.11 | 0.22 | 0.0406 | 0.0609 | 0.0864 |
| | Finetuned + H.I | 0.0662 | 0.07 | 0.07 | 0.37 | 0.0384 | 0.0384 | 0.1053 |
| | Finetuned + H.I + Dial. | 0.0702 | 0.07 | 0.13 | 0.28 | 0.0539 | 0.0729 | 0.1107 |
| TigerLLM | H.I Only | 0.0206 | 0.02 | 0.03 | 0.19 | 0.0143 | 0.0175 | 0.0517 |
| | H.I + Dial. | 0.0417 | 0.05 | 0.11 | 0.22 | 0.0352 | 0.0530 | 0.0770 |
| | Finetuned + H.I | 0.0776 | 0.07 | 0.15 | 0.24 | 0.0643 | 0.0896 | 0.1104 |
| | Finetuned + H.I + Dial. | 0.0911 | 0.10 | 0.12 | 0.19 | 0.0889 | 0.0952 | 0.1205 |
| Bangla LLaMA | H.I Only | 0.0229 | 0.02 | 0.03 | 0.05 | 0.0200 | 0.0232 | 0.0286 |
| | H.I + Dial. | 0.0309 | 0.03 | 0.05 | 0.08 | 0.0318 | 0.0361 | 0.0314 |
| | Finetuned + H.I | 0.0306 | 0.02 | 0.03 | 0.19 | 0.0143 | 0.0175 | 0.0517 |
| | Finetuned + H.I + Dial. | 0.0401 | 0.05 | 0.08 | 0.21 | 0.0319 | 0.0417 | 0.0690 |

## A2 Open-Vocabulary Evaluation Results

Table A5 reports the complete results under the open-vocabulary evaluation protocol. Unlike catalog-based evaluation, this setting allows models to generate recommendations without selecting exclusively from a predefined candidate catalog, thereby providing a complementary and more challenging evaluation of recommendation generation.

### A2.1 Template Diversity Calculation

We quantify the structural diversity of the 100 dialogue templates using normalized Shannon entropy, sequence uniqueness, and pairwise normalized edit distance. For a categorical template attribute $X$ with $K$ observed categories, normalized Shannon entropy is calculated as

$$H_{\text{norm}}(X) = \frac{-\sum_{i=1}^{K} p_i \log p_i}{\log K}, \tag{1}$$

where $p_i = n_i/N$, $n_i$ is the number of templates in category $i$, and $N = 100$. The normalized entropy ranges from 0 to 1, with higher values indicating a more balanced distribution across categories.

For recommendation count, the templates consist of 34 one-recommendation, 33 two-recommendation, and 33 three-recommendation templates. Because these categories are shown separately in Table A6, we report the normalized entropy contribution of each category as

$$h_i = \frac{-p_i \log p_i}{\log 3}. \tag{2}$$

Table A6: Structural diversity of the 100 dialogue templates. The recommendation-category scores are normalized entropy contributions and sum to an overall recommendation-count entropy of $0.9999$. All other diversity scores range from 0 to 1, with higher values indicating greater diversity.

| Measure | Count/Range | Diversity Score |
|---|---|---|
| One-recommendation templates | 34 | 0.334 |
| Two-recommendation templates | 33 | 0.333 |
| Three-recommendation templates | 33 | 0.333 |
| Dialogue-length distribution | 6--16 turns | 0.925 |
| Unique goal sequences | 84/100 | 0.988 entropy |
| Unique subgoal sequences | 100/100 | 1.000 entropy |
| Unique speaker sequences | 45/100 | 0.935 entropy |
| Goal-sequence variation | 4,950 pairs | 0.444 edit distance |
| Subgoal-sequence variation | 4,950 pairs | 0.722 edit distance |
| Speaker-sequence variation | 4,950 pairs | 0.266 edit distance |

The resulting contributions are $0.334$, $0.333$, and $0.333$, respectively, and sum to an overall normalized recommendation-count entropy of $0.9999$. Dialogue length contains 11 observed categories spanning 6--16 turns, with frequencies $\{4, 12, 15, 16, 11, 13, 12, 8, 6, 1, 2\}$. Applying Equation 1 gives a dialogue-length diversity score of $0.925$.

For goal, subgoal, and speaker structures, each template is represented as an ordered sequence of its corresponding turn-level labels. We report both the number of unique sequences and their normalized Shannon entropy. Structural differences between templates are additionally measured using normalized Levenshtein distance:

$$d_{\text{norm}}(s_i, s_j) = \frac{\text{Lev}(s_i, s_j)}{\max\left(|s_i|, |s_j|\right)}, \tag{3}$$

where $\text{Lev}(s_i, s_j)$ is the minimum number of insertions, deletions, and substitutions required to transform sequence $s_i$ into sequence $s_j$. We calculate this value for all

$$\binom{100}{2} = 4{,}950 \tag{4}$$

unique template pairs and report the mean pairwise distance. Values near 0 indicate highly similar structures, whereas values near 1 indicate greater structural variation.

## A3 Factual Grounding Evaluation

We evaluate whether claims in the generated dialogues are supported by the corresponding user ratings, reviews, and preference histories. The evaluation includes two native Bangla-speaking annotators, GPT-5.1 as an independent judge, and a shuffled-user control.

### A3.1 Human Annotation Guidelines

For each instance, annotators are shown the generated dialogue, a highlighted claim, the relevant product, user rating, source review, and Like/Dislike label. They evaluate only the highlighted claim using the supplied evidence and assign a score from 0 to 4:

- **4---Fully supported:** The claim is completely supported by the source record.

- **3---Mostly supported:** The claim is supported but contains a minor inference or exaggeration.
- **2---Partially supported:** Only part of the claim is supported.
- **1---Weakly related:** The claim is related to the product but has limited support.
- **0---Unsupported or contradicted:** The claim is absent from or conflicts with the source record.

Annotators are instructed not to use external knowledge, to distinguish missing evidence from contradiction, and to verify product identity and preference polarity. Faithful paraphrases are accepted, while unsupported attributes receive lower scores.

### A3.2 Shuffled-User Control

Each claim is evaluated against both the correct user record and a randomly selected user record from the same product domain. The conditions are presented in random order without revealing their identity. We calculate:

$$\Delta_{\text{grounding}} = S_{\text{correct}} - S_{\text{shuffled}}.$$

The correct-user condition obtains an average grounding score of 3.07, compared with 2.85 for the shuffled-user condition. The two annotators achieve Cohen's $\kappa = 0.65$, indicating substantial agreement.

### A3.3 LLM-Based Evaluation

GPT-5.1 evaluates the same claims using the same 0--4 scale and obtains an average grounding score of 3.23. The exact prompt is provided below.

```
You are evaluating the factual grounding of a claim from a
Bangla conversational recommendation dialogue.

Use only the supplied source record. Do not use external
knowledge or assume that plausible information is factual.

SOURCE USER RECORD

Product: {PRODUCT_NAME}
Category: {PRODUCT_CATEGORY}
Rating: {RATING}
Preference: {LIKE_OR_DISLIKE}
Review: {SOURCE_REVIEW}

GENERATED DIALOGUE

{GENERATED_DIALOGUE}

CLAIM TO EVALUATE

{HIGHLIGHTED_CLAIM}

Score the claim using the following scale:

4 = Fully supported
3 = Mostly supported
2 = Partially supported
1 = Weakly related
0 = Unsupported or contradicted

Also determine whether the claim directly contradicts the
review, rating, or preference label. Missing evidence alone
should not be treated as a contradiction.

Return only valid JSON:
```

```
{
  "grounding_score": 0,
  "contradiction": false,
  "evidence": "Supporting evidence or None",
  "unsupported_content": "Unsupported content or None",
  "rationale": "Brief explanation"
}
```

For the shuffled-user condition, we use the same prompt with a randomly selected user's source record while keeping the generated dialogue and highlighted claim unchanged.